\documentclass[aps,prb,floatfix,twocolumn]{revtex4}
\usepackage{graphicx}
\usepackage{amsmath}
\usepackage{amssymb}

\newcommand{\N}{{\cal N}}

\newcommand{\BEQ}{\begin{eqnarray}}
\newcommand{\EEQ}{\end{eqnarray}}
\newcommand{\BEA}{\begin{eqnarray}}
\newcommand{\EEA}{\end{eqnarray}}

\renewcommand{\d}{{\rm d}}

\newcommand{\tauc}{{\tau_{\rm c}}}
\newcommand{\taur}{{\tau_{\rm reg}}}

\newcommand{\tr}{{\rm tr}}

\newcommand{\HH}{\hat{H}}

\newcommand{\RS}{{\rm S}}
\newcommand{\RA}{{\rm A}}
\newcommand{\RM}{{\rm M}}
\newcommand{\RB}{{\rm B}} 
\newcommand{\RSA}{{\rm SA}}
\newcommand{\down}{{\downarrow}}
\newcommand{\up}{{\uparrow}}
\newcommand{\uu}{{\uparrow\uparrow}}
\newcommand{\dd}{{\downarrow\downarrow}}
\newcommand{\ud}{{\uparrow\downarrow}}
\newcommand{\du}{{\downarrow\uparrow}}

\newcommand{\ri}{{\rm i}}
\newcommand{\tf}{t_{\rm f}}

\newcommand{\um}{{\underline m}}

\newcommand{\Tc}{{T_{\rm c}}}
\newcommand{\half}{\frac{1}{2}}

\newcommand{\CD}{{\cal D}}
\begin{document} 
\title
{The quantum measurement process in an exactly solvable model}
\author{Armen E. Allahverdyan$^{1,2)}$, Roger Balian$^{3)}$
and \underline{Theo M. Nieuwenhuizen}$^{1)}$\footnote{The author who 
presented this contribution at
`Foundations of Probability and Physics-3', V\"axj\"o, Sweden,
June 7-12, 2004}}
\affiliation{$^{1)}$ Institute for Theoretical Physics,
Valckenierstraat 65, 1018 XE Amsterdam, The Netherlands}
\affiliation{$^{2)}$Yerevan Physics Institute,
Alikhanian Brothers St. 2, Yerevan 375036, Armenia}
\affiliation{$^{3)}$ SPhT, CEA-Saclay, 91191 Gif-sur-Yvette cedex, France}

\begin{abstract}
An exactly solvable model for a quantum measurement is discussed 
which is governed by hamiltonian quantum dynamics. 
The $z$-component $\hat s_z$ of a spin -$\half$ is measured with an apparatus, 
which itself consists of magnet coupled to a bath.
The initial state of the magnet is a metastable paramagnet, while the bath
starts in a thermal, gibbsian state. 
Conditions are such that the act of measurement drives the magnet in the 
up or down ferromagnetic state according to the sign of  $s_z$ of 
the tested spin. The quantum measurement goes in two steps.
On a timescale $1/\sqrt{N}$ the off-diagonal elements of the spin's density matrix
vanish due to a unitary
evolution of the tested spin and the $N$ apparatus spins; on a larger but
still short timescale this is made definite by the bath.
Then the system is in a `classical' state,
having a diagonal density matrix. The registration of that state is
a quantum process which can already be understood 
from classical statistical mechanics.
The von Neumann collapse and the Born rule are derived rather than postulated.

\end{abstract}

\maketitle

It is astonishing that after one century of success of the quantum description 
of nature, its foundations are as mysterious as ever
~\cite{wh},~\cite{Vaxjo},~\cite{deMuynck},~\cite{Schlosshauer}. To determine
the precise meaning of a wavefunction (or, more generally, a density matrix),
a fundamental understanding of the quantum measurement process is required, since
this is the only point of contact between theory and experiment.

To investigate the matter, several models have been 
proposed~\cite{models}, which did not converge to a unique picture.
Here we discuss an exactly solvable model~\cite{ABNEPL}, 
which retains all properties of realistic measurements and 
from which the general structure, found before in a more
complicated bosonic model~\cite{ABNBose}, can be read off. 

As foreseen on general grounds~\cite{vKampen},~\cite{balian},
the measurement appears to take place in two steps: 
on a quantum timescale $\tauc\ll \hbar/T $, 
disappearence of off-diagonal elements of the spin's density matrix
 occurs (vanishing of Schr\"odinger cat terms),
  while on a timescale $\taur\gg \hbar/T$ the registration 
of the measurement occurs. 
The registration is analogous to the measurement of an
ensemble of `classical' Ising spins $s_z$ taking the values $+1$ or $-1$.

As we shall discuss in the conclusion, our solution for
the measurement problem is compatible with the 
statistical interpretation of quantum  mechanics, 
as a theory that describes ensembles. It rules out several 
competing interpretations.

\subsection*{\it Classical measurement} 

The analysis of the quantum measurement, to be discussed below,
appears to exhibit some classical features. For this reason, and also for
its own sake, we first explain how to measure a classical 
Ising spin (classical two-state system), which is in a definite state 
$s_z=\pm 1$. It is known that some classical systems may indeed
be approximately described as a two-state object, e.g. a classical 
brownian particle in a double-well potential with well-separated minima 
and a steep potential barrier in between. 

Our apparatus (A) consists of a magnet (M) coupled to a phonon  bath (B). 
The magnet contains N Ising spins $\sigma_z^{(n)}=\pm 1$ having a 
mean-field interaction between all quartets 

$$H_\RM = -\frac{J}{4 N^3} \sum_{ijkl=1}^N 
\sigma_z^{(i)}\sigma_z^{(j)}\sigma_z^{(k)}\sigma_z^{(l)}
= -\frac{1}{4}NJ{\underline m}^4,\eqno(1)$$         
with ${\underline m}=(1/N)\sum_n\sigma_z^{(n)}$ denoting the fluctuating
magnetization. In the standard Curie-Weiss model all pairs would be
coupled, and a second order phase transition occurs.
The quartic interaction has been chosen in order to have a first order
transition, as it happens in a bubble chamber, where an oversaturated
liquid creates droplets of its stable phase, the gas,
when triggered by a particle.
In general, the apparatus should amplify the microscopic signal
and go to a stable pointer state so as to allow reading at an arbitrary moment. 
These conditions can indeed be met when  it
starts in a metastable state.

The interaction between the tested system S and the apparatus
$$ H_\RSA=-gs_z\sum_n\sigma_z^{(n)}=-gs_zN{\underline m},\eqno (2)$$
is turned on at $t=0$, the beginning of the measurement, and turned off
at, say, $\tf/2$, after which the apparatus is left to relax untill the 
final time $t_{\rm f}$.

Initially the magnet starts in the paramagnetic state: each spin has 
chance $\half$ to be up or down, implying a vanishing average magnetization,
$m\equiv\langle\um\rangle=0$.

At a critical temperature $\Tc$ the magnet undergoes  a phase transition
to one among two states with magnetization $m_\up>0$ and $m_\down=-m_\up<0$. 
Due to the quartic interactions (2), it is a first order phase transition.
The free energy, $F=U-TS$, is simply derived, owing to the fact that 
in this model the mean field approximation becomes exact for large $N$. 
The energy being obvious,  one needs the entropy $S=\log\Omega$. 
Since the degeneracy of states with magnetization $m$ 
equals $\Omega=N!/[(N_+)!\,\,\,(N_-)!]$, where $N_\pm=\half(1\pm m)N$, 
one gets immediately

$$ \frac{F}{N}=- \frac{J m^4}{4}-g s_zm-
T(\frac{1+m}{2}\ln\frac{2}{1+m}+\frac{1-m}{2}\ln\frac{2}{1-m}),
\eqno (3) $$ 
At $g=0$ and for $T$ below $\Tc=0.362949\,J$, the paramagnet $m=0$ is 
still metastable, see  Fig. 1. 
It is here that the setup lends itself as an apparatus for a measurement:
by starting in the metastable paramagnet this constitutes the magnetic
analogue of the metastable oversaturated liquid of a bubble chamber.

\begin{figure}
\includegraphics[width=7cm,height=5cm]{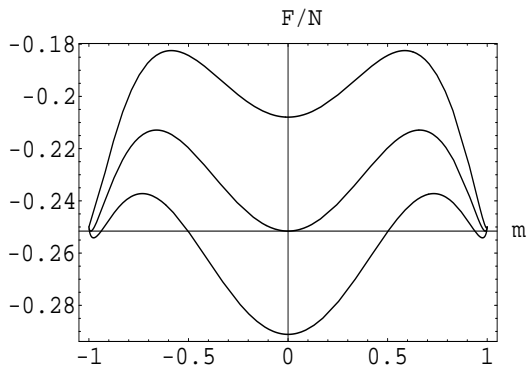}
\caption{Free energy of the magnet as function of $m$.
Lower curve: at large $T=0.42\,J$ the paramagnet $m=0$ has lowest free energy.
Middle curve: at $\Tc=0.362949\,J$ the local minima 
$m=0$, $m_\up=0.990611$ and $m_\down=-m_\up$ become degenerate.
Upper curve: Below $\Tc$, here  $T=0.3\,J$, the paramagnet is metastable, 
while the minima $m_\up$, $m_\down$ are stable. 
In the measurement the magnet starts in the metastable state 
and ends up in one of the stable states.}
\end{figure}

\begin{figure}
\includegraphics[width=7cm,height=5cm]{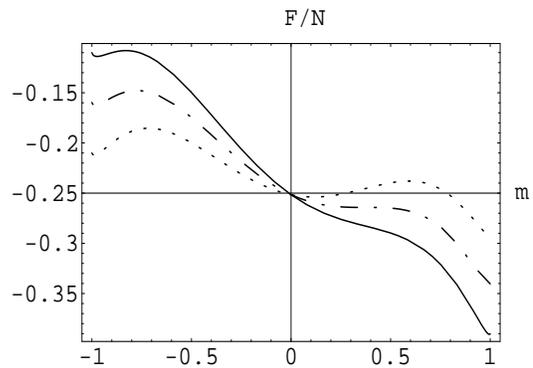}
\caption{A free energy barrier can be overcome 
by the coupling. Here $T=\Tc$ and $s_z=+1$. 
Dotted curve: the small coupling $g=0.04\,J$ does not suppress the barriers.
The setup cannot bring the magnetization from $m=0$ to the minimum near $m=1$. 
Dash-dotted curve: at the critical value $g_c= 0.09035\,J$  the barrier 
near $m=0.5$ is just suppressed. 
Full curve: at large coupling, $g=0.12\,J$, there is no barrier
and $m$ will end up in the minimum to register the measurement.
For $s_z=-1$ the left barrier would be suppressed.}
\end{figure}

At time $t=0^+$ the coupling $g$ between the tested spin and the apparatus
 is turned on, which puts the magnet in an external field $gs_z=\pm g$, see Eq. (2). 
If $g$ is large enough and $s_z=+1$, the interaction suppresses the barrier 
near $m=0.6$, see Figure 2, while for $s_z=-1$ it will suppress the one near $m=-0.6$.
This is the magnetic analogue of a bubble in a bubble chamber, where a supercritical 
liquid is triggered to bubbles of its gas state by a tested particle.

Let us denote by $\up$ and $\down$ the $s_z=\pm1$ cases. With the field 
turned on, the magnetization will move from $m=0$ to the
minimum of $F$. This is possible due to a weak coupling to the bath,
which allows dumping of the excess energy in the bath. 
In a classical approach one avoids going into details of the bath
by assuming a Glauber-type of dynamics for the spins of the magnet. 
Using this, the dynamics of $m$ has been coined on the basis 
of detailed balance alone~\cite{SuzKubo}, but that is not enough to fix it.
In a proper quantum mechanical treatment, one may 
consider the model where all three spin components of all $N$ apparatus spins
are weakly coupled to independent Ohmic bosonic subbaths 
(sets of harmonic oscillators). The proper dynamics then appears to be: 
$$\dot{m}=
\gamma h (1-\frac{m}{\tanh\beta h}),\quad h=gms_z+Jm^3, \eqno (4)
$$
where $\gamma\ll 1/\hbar$ is a small parameter charactering the 
weak coupling to the bath.  For $s_z=+1$,  $m$ will go to the right, see Fig. (2).
When $m$ has approached the minimum $+m_\ast\approx 0.994$, it remains stably close to 
$1$ whether S and A are coupled or not. 
After decoupling the apparatus ($g\to 0$), $m$ moves slightly from $m^\ast$ to  
the $g=0$ - minimum  $m_\up\approx 0.9906$. 
It will stay there up to a hopping time $\sim\exp(N)$; 
for large $N$ this means ``for ever''. Whether or not the apparatus is 
read off at any time (``observation'') is obviously of no significance for the 
measurement.  

The measurement has now been performed: if $s_z$ was $+1$, the apparatus has 
ended up with magnetization per spin $m_\up\approx 1$, and for $s_z=-1$ the 
magnetization per spin went to $m_\down=-m_\up$, so the sign of the tested spin 
is amplified in the macroscopic magnetization $Nm_\up$ or $Nm_\down$.

If there is an ensemble of spins, 
repeating the measurement will allow determination of the fraction
$p_\up=\N_\up/(\N_\up+\N_\down)$ of up-spins and the fraction
$p_\up=\N_\down/(\N_\up+\N_\down)$ of down-spins, where 
$\N_\up$ and $\N_\down$ are the number of measurements with magnetization
up and down, respectively.

\subsubsection*{\it Quantum measurement.} 

The above classical setup carries over immediately to the quantum 
situation. 
First, the Ising tested spin $s_z$ should be replaced by the 2x2 Pauli matrix 
$\hat s_z$, and the apparatus spins $\sigma_z^{(n)}$ by  $\hat \sigma_z^{(n)}$.
The magnetization operator $\hat m=(1/N)\sum_n \hat \sigma_z^{(n)}$ will
enter the Hamiltonians (1) and (3). In the Hamiltonian of the bath
and the interaction Hamiltonian between the magnet spins and the bath,
there will occur creation and annihilation operators for the bosons,
while the interaction term involves, apart from  those bosons,
the Pauli operators for the spins.

The tested spin may start in an unknown quantum state, that is to say, its
 spin-averages $\langle\hat s_x\rangle$, $\langle\hat s_y\rangle$ and 
$\langle\hat s_z\rangle$  are unknown and arbitrary. 
The measurement is expected to determine $\langle\hat s_z\rangle$, while the
information about  $\langle\hat s_{x,y}\rangle$ is expected to get lost. 
On the basis where $\hat s_z$ is diagonal 
the initial density matrix $\hat r(0)$ has the elements 
$r_{\uu}(0)=\half(1+\langle\hat s_z\rangle)$,   
$r_\ud(0)=\half(\langle\hat s_x\rangle-\ri\langle\hat 
s_y\rangle)=r_\du^\ast(0)$, $r_\dd(0)=\half(1-\langle\hat s_z\rangle)$.

The density matrix of the total system $\hat \CD$ is initially chosen as
a tensor product, $\hat\CD(0)=\hat r(0)\otimes \hat R_\RA(0)$,
 to express that there are initially no corrections between system and apparatus,
in order to avoid any bias in the measurement. 
The apparatus itself also has two uncorrelated parts, magnet and bath, viz.
$\hat R_\RA(0)=\hat R_\RM(0)\otimes \hat R_\RB(0)$, where 
$\hat R_\RM(0)$ describes the paramagnet, 
where each spin is independently up or down with chance $\half$, i.e.
 $\hat R_\RM(0)=2^{-N}\Pi_n \hat\sigma_0^{(n)}$ with 
the identity matrix $(\hat\sigma_0^{(n)})_{ij}=\delta_{ij}$,
while $\hat R_\RB(0)$ is the equilibrium (Gibbs) state of the bath.

\subsubsection*{\it Selection of the {\it collapse} basis}

The dynamics is set by the von Neumann equation 
$\ri \hbar\frac{\d}{\d t} \hat\CD=[\HH,\hat\CD]$,
where $\HH$ is the full Hamiltonian operator, including also the bath and 
the coupling between magnet and bath.
The state of the tested spin is $\hat r(t)=\tr _{\RM,\RB} \hat\CD(t)$.
For its evolution the quantum version of the 
interaction Hamiltonian (3) provides, 
$$ \frac{\d}{\d t}\, r_{ij}  
= -gN(s_i-s_j)\,\tr_{\RM,\RB}[\,\hat{m},\hat\CD_{ij} ], \eqno (5)$$
where $i,j=\up,\down$ and $s_\up=+1$,  $s_\down=-1$ are the eigenvalues 
of $\hat s_z$ and where the four blocks $\hat\CD_{ij}$ of $\hat\CD$ 
act in the apparatus space.
It follows that the diagonal elements are conserved in time: $r_\uu(t)=r_\uu(0)$,
$r_\dd(t)=r_\dd(0)$. This happens because the spin has no dynamics of its own.
The conservation is a sine-qua-non condition for a reliable ideal measurement.
The off-diagonal elements are not conserved since for them $s_i-s_j\neq 0$, 
so they are endangered and they will actually  vanish.

We learn from this argument that the selection of the  collapse basis is a direct 
consequence of the forces exerted by the apparatus on the test system:
The choice of the interaction Hamiltonian sets the basis 
on which the density matrix of the system diagonalizes. Zurek has claimed that the 
selection would be imposed by the coupling to the environment~\cite{Zurek},
even though the difficulties to control these couplings make it an 
undesired candidate for such an important issue. The above argument
does in no way invoke the environment and thus rules out Zurek's picture.

\subsubsection*{\it Disappearance of Schr\"odinger cats} 
 
At $t=0$ the coupling between system and apparatus is turned on.
Let us consider  very early times, where both the spin-spin 
interactions and the spin-bath interactions are still ineffective. 
The problem is then simply the evolution of $N$ independent 
apparatus spins, not coupled to the bath, in a field arising from the tested spin. 
This means that within $\hat\CD_\ud$ 
the density matrix of each spin evolves as
$ \hat\sigma_0^{(n)}$ = diag ($1,1$) $\to$
diag ($e^{2\ri gt/\hbar}$, $e^{-2\ri gt/\hbar}$), implying

$$r_\ud(t)\equiv \tr_{\RM,\RB}\hat \CD_\ud(t)
=r_\ud(0)[\cos\frac{2gt}{\hbar}]^N. \eqno(6)$$

For short times this produces  a gaussian decay,
$r_\ud(0)\,\exp(-t^2/\tau^2_{\rm c})$, with `cat' time

$$\tauc=\frac{\hbar}{g\sqrt{2N}}\ll \frac{\hbar}{T}. \eqno (7)$$
In the estimate we took $g\sim J\sim T$ and $N\gg 1$.

The matrix element (6) still presents recurrence peaks at $t_k=k\pi\hbar/2g$. 
However, provided $N$ is large they will be suppressed by the  bath, 
as it brings in a factor $\sim \exp(-\gamma\hbar N)$.
A small dispersion in the $g$'s is quite realistic, and it also brings 
a reduction $\exp(-k^2\pi^2\frac{\langle g^2\rangle-\langle g\rangle^2}
{2\langle g\rangle^2}\,N)$.

Altogether, in the quantum coherent process which takes place on the 
shortest time scale (7), the Schr\"odinger cat hides itself if $N$ is macroscopic, 
in agreement with von Neumann's postulate. 
The recurrence of the peaks is suppressed somewhat later, 
also on a short time scale, owing either to 
a small coupling of the macroscopic magnet with the bath 
or to a dispersion in the couplings $g$. In the first case the erasure 
of recurrences is an effect of the `environment',  but, contrary to 
what is often thought, see  Zurek ~\cite{Zurek} for a recent review,
the environment (bath) appears not to be the main cause.

\subsubsection*{\it Registration of the quantum measurement.} 

Once the off-diagonal sectors of the density matrix have decayed, 
there remain the diagonal ones, which evolve more slowly because 
of dumping energy in the bath. 
Now the bath, which we describe by a model simulating phonons, 
and the coupling to it have to be  specified in detail. 
For simplicity each spin of the magnet is assumed to have its own subbath. 
These subbaths are all identical but independent, consisting of 
harmonic oscillators in $x$, $y$ and $z$-direction, 
which are coupled bilinearly to the components
of the spins, and start out in their Gibbs state. The characteristic
coupling constant with the Ohmic bath is $\gamma$, and weak coupling means
that $\gamma \ll 1/\hbar$.
Working out this quantum problem we observe complete analogy to the above
description termed ``classical measurement". In particular, the evolution 
of $m(t)=\tr\,\hat m\hat\CD(t)$ is found to be given by Eq. (4) announced above. 
The characteristic timescale is much larger than the `cat' time,
$$\taur=\frac{1}{\gamma g}\sim\frac{1}{\gamma J}\sim\frac{1}{\gamma T}\gg
 \frac{\hbar}{T}.
\eqno (8)$$
The physical reason is that an extensive amount of energy has to be
transfered to the bath; this takes a time $\sim 1/\gamma$ since 
the characteristic coupling constant with the bath is $\gamma$.
The stable points of the dynamics are the minima of the free energy 
discussed above.

The last stage, where after decoupling the apparatus (by setting $g=0$)
$m$ is stabilized at either $m_\up$ or $m_\down$, proceeds as in the 
classical case.

In short, registration of the quantum measurement is the same as for 
the above classical measurement. This might perhaps have been anticipated from 
the fact that on the considered timescale the diagonalization
has already taken place, so the density matrix is ``classical".

\subsubsection*{\it Post-measurement state}

After the measurement, at $\tf\gg \taur$, the common state of the tested spin 
and apparatus is stationary and equal to
$$ \label{Dfin3}
\hat\CD(\tf)=r_{\up\up}(0)|\up\rangle\langle \up| \otimes \hat R_{\RA\up}(\tf)
+r_{\down\down}(0)|\down\rangle\langle\down|\otimes \hat R_{\RA\down}(\tf)$$ 
\vspace{-1cm}
$$\hfill\hfill\eqno(9) $$
where $\hat R_{\RA\up}(\tf)$ is the product of a gibbsian state for the bath
and of the state
$$
\hat R_{\RM\up}(\tf)=\hat \rho_{\up\up}^{(1)}(\tf)\otimes \cdots \otimes 
\hat \rho_{\up\up}^{(N)}(\tf)  \eqno(10) $$
for the magnet, and where 
where $\hat\rho_\uu^{(n)}(\tf)=\half\, {\rm diag}(1+m_\up,1-m_\up)$ 
is the Gibbs density matrix of spin $n$ for the magnet, and where 
in the down sector one replaces $m_\up\to m_\down=-m_\up$.

The off-diagonal sectors, called ``Schr\"odinger cat components", 
have been eliminated by the initial evolution. Strictly speaking,
 the final state 
must be unitarily related to the initial state. However, the entropy of (9) 
is larger than the initial entropy. 
The solution of this paradox is the same as 
the solution of the paradox of irreversiblilty in statistical mechanics: 
Eq. (9) has been derived by using some approximations and relates to a
coarse grained entropy, but it differs from 
the exact state only through non-observable terms. The latter involve correlations 
of many degrees of freedom of the apparatus which for large $N$ 
become negligible for almost all observables 
except for the fine-grained entropy, which is conserved but of no relevance.

\subsubsection*{Born rule from statistics of pointer variables}

From the result (9) for the final quantum state of the system and the apparatus 
we can derive the characteristic function and hence the joint probability distribution 
for the pointer variable and the $z$-component of the tested spin. 
This distribution is peaked for large $N$ at two points: The average magnetization 
per spin $m$ can take the two values $m_\up$ or $m_\down$ with respective 
probabilities $p_\up = r_\uu(0)$ or $p_\down = r_\dd(0)$, and these two occurrences 
are completely correlated with the final state up or state down of the tested system.

This means that, when an ensemble of measurements is performed, two outcomes are possible
for each event. What is observed is the pointer variable $m$ of the apparatus after the process. 
From (9) we can calculate the moments of magnetization per spin after the measurement,
$
\langle m^k\rangle(\tf)\equiv\tr\,\hat m^k\hat\CD(\tf)=r_{\up\up}(0)m_\up^k
+r_{\down\down}(0)m_\down^k, $
which just confirms the dichotomic distribution of $m(\tf)$. We may indeed
consider $P(m;\tf)=\tr_{\RS,\RA}\delta(\hat m-m)\hat\CD(\tf)$,  which exposes
the dichotomicity most directly,
$$
P(m;\tf)=r_{\up\up}(0)\delta(m-m_\up)+r_{\down\down}(0)\delta(m-m_\down). \eqno(11)
$$
Quantum mechanics confronts us with this macroscopic relation; its application to nature
should set its interpretation.  It is from
(11), which refers to observations about the directly observable quantity $m$,
and moreover from the theory which shows that (9) encompasses full classical
correlations between $m$ and $s_z$, that
interpretations of quantum mechanics can be confronted with experiments.
 But this is rather standard when we realize that
(11) is a relation referring to the macroscopic world only. 
From experimental practice we know that what is observed is the pointer
variable of the apparatus after 
an individual measurement, the final magnetization, which equals $Nm_\up$ or $Nm_\down$. 
Classical probability theory then says that we are dealing with an 
{\it ensemble of measurements on an ensemble of systems or system preparations}, 
and that the factor $r_{\up\up}(0)$ in (11) should  be identified with the fraction 
$p_\up=\N_\up/(\N_\up+\N_\down)$ individual measurements (`events')  
where a positive magnetization $Nm_\up$ is observed.

In each of these occurrences we can infer indirectly
from quantum theory, relying on (9), that
the tested system is prepared in the pure state with $s_z=+1$
(this is sometimes called  collapse of the wavefunction or reduction of the
wavepacket). This system-apparatus connection becomes more practical if, 
analogous to (11), we maintain about the apparatus only the information 
regarding its pointer variable. This is done by considering
$\hat R(m;\tf)=\tr_\RA\,\delta(\hat m-m)\, \hat\CD(\tf)$, which is 
a classical distribution function of the pointer variable
and an operator in the Hilbert space of the tested system. 
Eq. (9) yields
$$
\hat R(m;\tf)=r_{\up\up}(0)\delta(m-m_\up)\,|\up\rangle\langle\up|
+r_{\down\down}(0)\delta(m-m_\down)|\down\rangle\langle\down|. \eqno(12)
$$
For practical applications this result of the measurement process 
defines a broad class of ideal measurements, whereas (9) describes the complete 
 final state of the apparatus, that, however, is  hardly ever tested.

    It is thus the theoretical analysis of the interaction process between
the system and the apparatus which allows us to regard this process as a
``measurement" in which microscopic quantum information is deduced from
macroscopic observation. In order to get this information on $s_z$ in the
form of an ordinary probability, we had to pay a tribute: the loss of the
initial information about the off-diagonal elements $r_{\up\down} (0)$ and 
$r_{\down\up} (0)$. The emergence of classical probabilities is due to the 
macroscopic size of the apparatus, and it is accompanied by a destruction 
of genuinely quantum elements. 

    Born's rule, together with von Neumann's collapse, are thus 
derived for large $N$ from the 
joint quantum evolution of the system and the apparatus.
The measurement process, accompanied by a sorting of the outcomes
of the pointer variable, can be used as a preparation of the system in a pure 
eigenstate of  $\hat s_z$.

\subsubsection*{ The statistical (ensemble) interpretation of quantum mechanics.}

In agreement with the preceding analysis, it is natural to describe 
the quantum measurement by adopting the statistical interpretation 
put forward by Einstein, see e.g. ~\cite{ballentine},
~\cite{balian}. The most important aspects are:
1) A quantum state is described by a density matrix.
2) A single system does not have ``its own" density matrix or wavefunction;
3) Each quantum state describes an ensemble of identically prepared systems;
this also  holds for a pure state $|\psi\rangle\langle\psi|$.

To give an example, in an ideal Stern-Gerlach experiment all particles in the upper
beam together are described by the ket $|\up\rangle$ or density matrix
$|\up\rangle\langle\up|$.

In this philosophy, a quantum measurement must describe  an ensemble of 
measurements on an ensemble of systems. 
This is indeed a natural interpretation of the post-measurement state (9).
 In doing a series of experiments, there are two possible
outcomes, connected with the magnetization of the apparatus being up or down,
which occur with probabilities $p_\up=r_\uu(0)$ and $p_\down=r_\dd(0)$, respectively.
In each such event, the $z$-component of the tested spin is 
equal to $+1$ (up) or $-1$ (down), correspondingly. 
The quantum subensemble of spins having $\up$ is described by
the pure state density matrix $|\up\rangle\langle \up|$, or simply
by the wavefunction $|\up\rangle$. A similar statement holds for the
down spins. This is von Neumann's  collapse postulate, and it arises 
here as a physical consequence of  quantum mechanics itself, 
taking into account that the apparatus is macroscopic.

Notice that the very same eq. (9) and its interpretation would arise from
the thermodynamics of measurements on an ensemble of classical Ising spins $\pm 1$,
thus merging classical and quantum measurements.
In the classical case one would be accustomed to trace out the bath, 
which would replace $\hat R_{\RA\up}(\tf)$ by $\hat R_{\RM\up}(\tf)$,
but we refrained from doing this, because of the confusion 
in the literature about its justification. 
We now see that also in the quantum situation
no information is lost when taking this trace,
because the off-diagonal terms $|\up\rangle\langle\down|$ and
$|\down\rangle\langle\up|$ have become inobservably small  anyhow. 

Let us notice that the statistical interpretation makes sense at all times,
at $t=0$, during the measurement and after the measurement.
Our mixed initial state of the apparatus describes a realistic preparation
of the apparatus, as opposed to the often assumed pure initial states, that
can in practice not be produced for any system with many degrees of freedom.
We thus do not consider pure state setups - that might have their own interest - 
as realistic measurement setups.

\subsubsection*{\it Comparison with von Neumann-Wigner measurement theory}

To compare our results with the standard description of the quantum measurement
~\cite{wh},~\cite{deMuynck},~\cite{Schlosshauer}
is not easy, because the latter does not embody the same physics and 
it is based more on assumptions than on derivations. 
Since it has, in our view, not solved the measurement problem,
the best we can do is to mention some analogies and differences.
In the von Neumann-Wigner approach one assumes that the apparatus starts 
in a pure state $|a_0\rangle$. 
Typically also the tested system is assumed to start in a pure state, 
$|\psi\rangle=c_\up|\up\rangle+c_\down|\down\rangle$. 
It is then assumed~\cite{deMuynck} that in the so-called {\it premeasurement} stage
$0<t<\tauc$, 
the total wavefunction develops as $|\Psi\rangle=|a_0\rangle|\psi\rangle\to 
|\Psi_{\rm c}\rangle=U|\Psi\rangle
=c_\up|\up\rangle|a_\up\rangle+c_\down|\down\rangle |a_\down\rangle$, where
the $|a_{\up,\down}\rangle$ are states of the apparatus assumed to be in 
one-to-one correspondence with its final pointer states at $\tf$.
The pure density operator $|\Psi_{\rm c}\rangle\langle\Psi_{\rm c}|$ differs
from our Eq. (9), but it has same final marginal state for the tested system, 
$\hat r(\tauc)=|c_\up|^2|\up\rangle\langle \up|+|c_\down|^2|\down\rangle\langle\down|$. 
The state for the apparatus 
$\hat R_\RA^{\rm vNW}(\tauc)=
|c_\up|^2|a_\up\rangle\langle a_\up|
+|c_\down|^2|a_\down\rangle\langle a_\down|$
would agree with the result obtained by tracing out the tested spin from eq. (9), 
if we would disregard the mixed nature of our states 
$\hat R_{\RA\,\up,\down}(\tauc)$. 
However, in order to interpret the process as a measurement, we need the off-diagonal terms 
in the overall state of S+A to dissapear. This is achieved in our model by the first stage of
 the evolution. Both this model and the 
von Neumann-Wigner theory involve unitary dynamics. However, in our treatment,
the off-diagonal elements $|\up\rangle\langle \down|$ 
and $|\down\rangle\langle \up|$  vanish very rapidly (more precisely, they lead to very 
small, non-observable terms), due to the mixed nature of the initial state of the apparatus,
while the bath or randomness of coupling is needed to make this erasure permanent. 
In the standard von Neumann Wigner approach, they survive because the initial state of the
 apparatus is pure; 
the question of how to discard  them has led to various interpretations of quantum mechanics. 
Here the disappearance of these terms appears simply as a statistical phenomenon, due to the 
averaging over the initial disordered state. Anyhow, a pure state of the apparatus is 
unrealistic on physical grounds.




A next claim of the   von Neumann-Wigner approach is that the statistics of pointer 
variables is correctly described. But it is well known that their marginal state 
can arise from all kinds of  sets of mixed non-orthogonal states of the form 
$\{|a_\alpha\rangle\langle a_\alpha|\}_{\alpha=1}^k$ with $k\geq 2$:
${\hat R}_\RA^{\rm vNW}(\tauc)$$
=\sum_{\alpha=1}^k\, \lambda_\alpha |a_\alpha\rangle\langle a_\alpha|$, 
and these states have in general nothing to do with spin up or spin down states along the 
$z$-axis. In other words, it is unclear why the chosen apparatus measures the
spin in the $z$-direction. This non-uniqueness of the measurement basis 
(`prescribed ensemble fallacy') is related to the fact that 
typically no interaction Hamiltonian between system and apparatus is
specified, while in our approach it was vital for the selection of this basis.

\subsubsection*{\it Other interpretations of quantum mechanics.}

Our results make some of the interpretations, that caused much dispute in
the past~\cite{wh},~\cite{Vaxjo},~\cite{Schlosshauer} obsolete.

In the standard Copenhagen interpretation, it is stated that ``the wave function
is the most complete description of the system", in other words, each closed 
system has its own wavefunction, a fact denied in the statistical interpretation. 
A criticism of our approach might be that 
``the problem is not solved because somehow there is the (unknown) wavefunction
of the total system", which cannot end up in a mixture.
This argument sets out, however, that both the system and the apparatus start
out in a pure state, which is unrealistic. We acknowledge, indeed, that it is 
impossible to prepare a macroscopic system in a pure state. 
Afterall, that would require a macroscopic number of post-measurement selections; 
what an experimentalist does is completely the opposite: 
he turns on the apparatus and waits until it has stabilized, after
which the measurement is carried out.  

The assumption of an underlying pure state for the whole system is 
unnecessary and would anyhow be problematic for describing 
the statistical nature of the apparatus in realistic setups.
Moreover, it would prevent the appearance of a domain with classical features.
 
Interestingly, our quantum mechanical description of the registration 
bears certain classical features, an issue imagined long ago by Bohr.

The multi-universe picture was devised to suppress the possibility of collapse, 
a phenomenon which seems to contradict the unitarity of the microscopic quantum
evolutions. We have seen, however, that the  diagonalization is a real process of 
quantum statistical physics, which occurs owing to the necessarily large size 
of the apparatus; collapse in individual events is the physical realization 
of this. Moreover, for the case of finite but large $N$, 
our model describes a good measurement up to a long but finite time, while the 
multi-universe picture is silent about this situation, the only one met in
practice.

Mind-body problems do not show up, because the act of observation is no more
than gathering information about the classical final state of the apparatus, 
which has already registered the relevant data. 
Whether or not one observes the outcome has no effect on the system. 
Observation appears just as a means for selecting a statistical subensemble 
with well defined spin, owing to the system-apparatus correlations.

We have shown that the  collapse is not caused by the environment~\cite{Zurek}, 
but  by the coupling to the apparatus which selects the  collapse basis. 

Gravitation, sometimes put forward, plays no role.

Extensions of quantum mechanics, like spontaneous or stochastic 
localization and spontaneous  collapse models,  are not needed.


We find no support for interpretations that attribute a special role 
to pure states of the tested spin, e.g. the modal interpretation
and bohmian or nelsonian mechanics.

\subsubsection*{\it Conclusion}

The initial paramagnetic state of the apparatus, which cannot be fully specified, 
consists of many microstates, so a statistical description is called for, and we have 
retained it to account for the quantum measurement. This is 
possible within the statistical interpretation of quantum mechanics, which 
states that any quantum state describes an ensemble of systems.
A theory of quantum measurements must therefore describe an ensemble of measurements 
on an ensemble of identically (fully or partially) prepared systems.

It was found that the off-diagonal terms (Schr\"odinger cat states) disappear quite fast 
after the start of the measurement. It goes in two steps: the
disappearence proper occurs due to interaction 
of the tested system with the macroscopic apparatus, and later is made definite 
either by bath-induced decoherence or already by randomness in the interaction.

The registration of the measurement occurs in a ``classical'' state,
a state that has already a diagonal density matrix. 
Here a naive classical approach and a detailed quantum 
approach yield the same outcome. The pointer variable ends up in a 
stable thermodynamic equilibrium state. 
Whether the outcome is observed or not is immaterial.

For a macroscopic apparatus disappearence of off-diagonal terms 
is almost instantaneous, yielding the basis for the postulate of von Neumann.
The Born rule follows from the statistics of pointer values.

Our theory can be tested by mapping out the $N$-dependence.

An important issue, namely whether single events can be accounted for by 
quantum mechanics, has to be answered negatively, since this would be incompatible 
with Eq. (11).  Within the statistical
interpretation, {\it quantum mechanics is a theory about the statistics of 
outcomes of experiments}, but it is unable to account for a single process. 
To describe single outcomes, a richer theory 
(``sub-quantum mechanics''  or ``hidden variable theory'') might be dreamt of.

{\it Acknowledgements}.
The work of A.E. A. is part of the research programme of the Stichting voor 
Fundamenteel Onderzoek der Materie (FOM), financially supported by 
the Nederlandse Organisatie voor Wetenschappelijk Onderzoek (NWO). 
Th.M. N. grateful for hospitality at V\"axj\"o University
and thanks Peter Keefe for a critical reading of the manuscript.

\end{document}